\documentclass[twocolumn,prd,superscriptaddress,showpacs,floatfix,%
preprintnumbers,nofootinbib]{revtex4}

\usepackage{epsfig}
\usepackage{bm}

\long\def\ignore#1{}

\usepackage[dvips]{color}
\definecolor{Black}{named}{Black}
\definecolor{Blue}{named}{Blue}
\definecolor{Red}{named}{Red}

\newcommand{\D}{{\rm d}}

\begin{document}

\title{Role of dense matter in collective supernova
neutrino transformations}

\author{A.~Esteban-Pretel}
\affiliation{Institut de F\'\i sica
 Corpuscular (CSIC-Universitat de Val\`encia),
 Edifici Instituts d'Investigaci\'o, Apt.\ 22085,
 46071 Val\`encia, Spain}

\author{A.~Mirizzi}
\affiliation{Max-Planck-Institut f\"ur Physik
(Werner-Heisenberg-Institut), F\"ohringer Ring 6, 80805 M\"unchen,
Germany}
\affiliation{Istituto Nazionale di Fisica Nucleare, Roma,  Italy}

\author{S.~Pastor}
\affiliation{Institut de F\'\i sica
 Corpuscular (CSIC-Universitat de Val\`encia),
 Edifici Instituts d'Investigaci\'o, Apt.\ 22085,
 46071 Val\`encia, Spain}

\author{R.~Tom\`as}
\affiliation{\hbox{II.~Institut f\"ur theoretische
Physik,~Universit\"at Hamburg,~Luruper
Chaussee~149,~22761~Hamburg,~Germany}}

\author{G.~G.~Raffelt}
\affiliation{Max-Planck-Institut f\"ur Physik
(Werner-Heisenberg-Institut), F\"ohringer Ring 6, 80805 M\"unchen,
Germany}

\author{P.~D.~Serpico}
\affiliation{Center for Particle Astrophysics, Fermi National
Accelerator Laboratory, Batavia, IL 60510-0500, USA}

\author{G.~Sigl}
\affiliation{\hbox{II.~Institut f\"ur theoretische
Physik,~Universit\"at Hamburg,~Luruper
Chaussee~149,~22761~Hamburg,~Germany}}
\affiliation{APC~\footnote{UMR 7164 (CNRS, Universit\'e Paris 7, CEA,
Observatoire de Paris)} (AstroParticules et Cosmologie), 10, rue
Alice Domon et L\'eonie Duquet, 75205 Paris Cedex 13, France}

\date{3 July 2008}

\preprint{FERMILAB-PUB-08-207-A, IFIC/08-34, MPP-2008-66}

\begin{abstract}
For neutrinos streaming from a supernova (SN) core, dense matter
suppresses self-induced flavor transformations if the electron
density $n_e$ significantly exceeds the neutrino density $n_\nu$ in
the conversion region. If $n_e$ is comparable to $n_\nu$ one finds
multi-angle decoherence, whereas the standard self-induced
transformation behavior requires that in the transformation region
$n_\nu$ is safely above~$n_e$. This condition need not be satisfied
in the early phase after SN core bounce. Our new multi-angle effect
is a subtle consequence of neutrinos traveling on different
trajectories when streaming from a source that is not point--like.
\end{abstract}

\pacs{14.60.Pq, 97.60.Bw}

\maketitle

\section{Introduction}                        \label{sec:introduction}

The neutrinos streaming from a supernova (SN) core or from the
accretion torus of coalescing neutron stars are so dense that the
neutrino-neutrino interaction causes collective flavor
transformations~\cite{Pantaleone:1992eq, Samuel:1993uw, Qian:1995ua,
Pastor:2002we, Sawyer:2005jk, Fuller:2005ae, Duan:2005cp, Duan:2006an,
Hannestad:2006nj, Balantekin:2006tg, Duan:2007mv, Raffelt:2007yz,
EstebanPretel:2007ec, Raffelt:2007cb, Raffelt:2007xt, Duan:2007fw,
Fogli:2007bk, Duan:2007bt, Duan:2007sh, Dasgupta:2008cd,
EstebanPretel:2007yq, Dasgupta07, Duan:2008za, Dasgupta:2008my,
Sawyer:2008zs, Duan:2008eb, Chakraborty:2008zp, Dasgupta:2008cu}.  At
the same time, the density of ordinary matter is also very large,
suppressing normal flavor conversions unless the neutrinos encounter
an MSW
resonance~\cite{Dighe:1999bi,Schirato:2002tg,Fogli:2003dw,Tomas:2004gr}.
On the other hand, one of the many surprises of self-induced flavor
transformations has been that dense matter barely affects them. They
are driven by an instability in flavor space that is insensitive to
matter because it affects all neutrino and antineutrino modes in the
same way. Therefore, it can be transformed away by going to a rotating
frame in flavor space~\cite{Duan:2005cp, Duan:2006an} in a sense to be
explained below. Various numerical simulations confirm this insight.

We here clarify, however, that the matter density can not be
arbitrarily large before it affects collective flavor conversions
after all. The matter term is ``achromatic'' only if we consider the
time--evolution of a homogeneous (but not necessarily isotropic)
neutrino ensemble on a homogeneous and isotropic matter background.
If the matter background is not isotropic, the current-current
nature of the neutrino-electron interaction already implies that
different neutrino modes experience a different matter effect.

It is more subtle that even without a current, matter still affects
different neutrino modes differently if we study neutrinos streaming
from a source. The relevant evolution is now the flavor variation of
a stationary neutrino flux as a function of distance. For a
spherically symmetric situation, ``distance from the source'' is
uniquely given by the radial coordinate $r$. Neutrinos reaching a
certain $r$ have travelled different distances on their trajectories
if they were emitted with different angles relative to the radial
direction. Therefore, at $r$ they have accrued different oscillation
phases even if they have the same vacuum oscillation frequency and
even if they have experienced the same matter background. In other
words, if we project the flavor evolution of different angular modes
on the radial direction, they have different effective vacuum
oscillation frequencies even if they have the same energy. The same
argument applies to matter that modifies the oscillation frequency
in the same way along each trajectory, but therefore acts
differently when expressed as an effective oscillation frequency
along the radial direction.

The neutrino-neutrino interaction, when it is sufficiently strong,
forces different modes to reach a certain $r$ with the same
oscillation phase. To achieve this ``self-main\-tained coherence''
the neutrino-neutrino term must overcome the phase dispersion that
would otherwise occur. Such a dispersion is caused not only by a
spectrum of energies, but also by a matter background.

To clarify and quantify these ideas we revisit in
Sec.~\ref{sec:homogeneous} the time evolution of a homogeneous
ensemble and introduce the notion of the ``rotation-averaged
equations of motion.'' In Sec.~\ref{sec:stream} we discuss the
geometric modifications when we study neutrinos streaming from a
source and quantify the multi-angle effect caused by ordinary
matter. We conclude in Sec.~\ref{sec:discussion} with a discussion
of the implications of our findings for realistic situations
relevant for supernova neutrinos.

\section{Homogeneous Ensemble}                 \label{sec:homogeneous}

\subsection{Isotropic background}

To understand the role of matter in collective neutrino
transformations we begin with the equations of motion (EOMs) in their
simplest form, relevant for a homogeneous (but not necessarily
isotropic) gas of neutrinos. We only consider two-flavor oscillations
where the most economical way to write the EOMs is in terms of the
usual flavor polarization vectors ${\bf P}_{\bf p}$ for each mode
${\bf p}$ and analogous vectors $\bar{\bf P}_{\bf p}$ for the
antineutrinos,
\begin{equation}
\dot{\bf P}_{\bf p}={\bf H}_{\bf p}\times{\bf P}_{\bf p}\,.
\end{equation}
The ``Hamiltonian'' is
\begin{equation}
 {\bf H}_{\bf p}=\omega_{\bf p}{\bf B}+\lambda{\bf L}+
 \mu\int\D{\bf q}\,
 \left({\bf P}_{\bf q}-\bar{\bf P}_{\bf q}\right)
 \left(1-{\bf v}_{\bf q}\cdot{\bf v}_{\bf p}\right)
\end{equation}
with $\D{\bf q}=\D^3{\bf q}/(2\pi)^3$. We have used the vacuum
oscillation frequency $\omega_{\bf p}=\Delta m^2/2E$ with $E=|{\bf
p}|$ for relativistic neutrinos, ${\bf B}$ is a unit vector in the
mass direction in flavor space, and ${\bf L}$ is a unit vector in the
weak-interaction direction with ${\bf B}\cdot{\bf L}=\cos2\theta$ and
$\theta$ the vacuum mixing angle. The effect of a homogeneous and
isotropic medium is parametrized by $\lambda=\sqrt2\,G_{\rm
F}(n_{e^-}-n_{e^+})$ whereas the neutrino-neutrino term is given by
$\mu=\sqrt2\,G_{\rm F}n_{\bar\nu_e}$. For simplicity we here assume
that initially only $\nu_e$ and $\bar\nu_e$ are present with an
excess neutrino density of
$n_{\nu_e}=(1+\epsilon)\,n_{\bar\nu_e}$~\cite{EstebanPretel:2007ec}.
The polarization vectors are 
initially
normalized such that 
$|\int\D{\bf p}\, \bar{\bf P}_{\bf p}|=1$
and
$|\int\D{\bf p}\, {\bf P}_{\bf p}|=1+\epsilon$. 
For antineutrinos the Hamiltonian is the same with
$\omega_{\bf p}\to-\omega_{\bf p}$.

The matter term is ``achromatic'' in that it affects all modes of
neutrinos and antineutrinos in the same way. It was first pointed out
by Duan et al.~\cite{Duan:2005cp} that therefore one may study the
EOMs in a coordinate system that rotates around ${\bf L}$ with
frequency $\lambda$ so that the matter term disappears. In the new
frame the vector ${\bf B}$ rotates fast around ${\bf L}$ so that its
transverse component averages to zero. Therefore, in the new frame
the rotation-averaged Hamiltonian is
\begin{equation}\label{eq:EOM2}
 \langle{\bf H}\rangle=\omega\cos(2\theta)\,{\bf L}+
 \mu({\bf P}-\bar{\bf P})\,,
\end{equation}
where for the moment we consider the even simpler case of an
isotropic and monochromatic neutrino ensemble where the entire system
is described by one polarization vector ${\bf P}$ for neutrinos and
one $\bar{\bf P}$ for antineutrinos.

A dense matter background effectively projects the EOMs on the
weak-interaction direction. In particular, the relevant vacuum
oscillation frequency is now $\omega\cos2\theta$. For a small mixing
angle, the case usually considered in this context, this projection
effect is not important. However, a large mixing angle would strongly
modify the projected $\omega$. Maximal mixing where $\cos2\theta=0$
would prevent any collective flavor transformations, an effect that
is easily verified in numerical examples.

One usually assumes that the (anti)neutrinos are prepared in
interaction eigenstates so that initially ${\bf P}$ and $\bar{\bf P}$
are oriented along ${\bf L}$. Therefore, the rotation-averaged EOMs
alone do not lead to an evolution. However, in the unstable case of
the inverted mass hierarchy, an infinitesimal disturbance is enough
to excite the transformation. The fast-rotating transverse ${\bf B}$
component that was left out from the EOM is enough to trigger the
evolution, but otherwise plays no crucial role~\cite{Duan:2005cp,
Hannestad:2006nj}.

If we consider a homogeneous system where $\mu$ is a slowly
decreasing function of time, one can find the adiabatic solution of
the EOM for the simple system consisting only of ${\bf P}$ and
$\bar{\bf P}$~\cite{Duan:2007mv, Raffelt:2007xt}. In vacuum, this is
a complicated function of $\epsilon$ and $\cos2\theta$. In dense
matter, however, we are effectively in the limit of a vanishing
mixing angle because the initial orientation of the polarization
vectors now coincides with the direction relevant for the
rotation-averaged evolution. The original vacuum mixing angle only
appears in the expression for the projected oscillation frequency
$\omega\cos2\theta$.

With $z=\bar P_z$ the adiabatic connection between $\bar P_z$ and
$\mu$ is now given by the inverse function of
\begin{equation}\label{eq:adiabatic}
 \frac{\omega\cos2\theta}{\mu}=\frac{\epsilon+2z}{2}
 -\frac{\epsilon+2z+(3\epsilon+2z)z}
 {2\sqrt{(1+z)(1+z+2\epsilon)}}\,,
\end{equation}
where $-1\leq z\leq+1$. The ``synchronization radius'' $r_{\rm sync}$
where the adiabatic curve begins its decrease is implied by $\bar
P_z=z=1$. One finds the familiar result~\cite{Hannestad:2006nj,
Duan:2007mv}\footnote{In these papers $\mu$ was normalized to the
density of neutrinos and the results were expressed in terms of
$\alpha=1/(1+\epsilon)$. Here we have normalized $\mu$ to the density
of antineutrinos and use the picture of an excess of neutrinos,
expressed by $\epsilon$. For the same physical system our $\mu$ is
the one of Ref.~\cite{Hannestad:2006nj}, divided by $1+\epsilon$.}
\begin{equation}\label{eq:synch}
\frac{\omega\cos2\theta}{\mu}\Big|_{\rm
sync}=\frac{(\sqrt{1+\epsilon}-1)^2}{2}\,.
\end{equation}
For $\mu$ values larger than this limit, the polarization vectors are
stuck to the ${\bf L}$ direction.

Without matter one finds that $({\bf P}-\bar{\bf P})\cdot{\bf B}$ is
conserved~\cite{Hannestad:2006nj}. Here, the analogous conservation
law applies to $({\bf P}-\bar{\bf P})\cdot{\bf L}$. Therefore, the
adiabatic solution for $P_z$ is such that $P_z-\bar P_z$ is
conserved, i.e., $P_z=\bar P_z+\epsilon$.

In summary, the presence of dense matter simplifies the EOMs and in
that the adiabatic solution is the one for a vanishing vacuum mixing
angle, provided one uses the projected vacuum oscillation frequency.

\subsection{Background flux}

As a next example we still consider a homogeneous system, but now
allow for a net flux of the background matter, assuming axial
symmetry around the direction defined by the flux. For simplicity we
consider a monochromatic ensemble with a single vacuum oscillation
frequency $\omega$. We characterize the angular neutrino modes by
their velocity component $v$ along the matter flux direction. The
EOMs are in this case
\begin{equation}
\dot{\bf P}_{v}={\bf H}_{v}\times{\bf P}_{v}\,.
\end{equation}
The Hamiltonian for the mode $v$ is
\begin{equation}
 {\bf H}_{v}=\omega{\bf B}+(\lambda-\lambda'v){\bf L}+
 \mu({\bf D}-v{\bf F})\,,
\end{equation}
where $\lambda'\equiv\lambda v_e$, $v_e$ being the net electron
velocity. Moreover,
\begin{eqnarray}
 {\bf D}&=&\int_{-1}^{+1}\D v\,
 \left({\bf P}_v-\bar{\bf P}_v\right)\,,
 \nonumber\\*
 {\bf F}&=&\int_{-1}^{+1}\D v\,v
 \left({\bf P}_v-\bar{\bf P}_v\right)
\end{eqnarray}
are the net neutrino density and flux polarization vectors. The
normalization is $\int \D v\,\bar{\bf P}_v=1$ and $\int \D v\,{\bf
P}_v=1+\epsilon$.

Next, we transform the EOMs to a frame rotating with frequency
$\lambda$, allowing us to remove the matter term, but not the matter
flux,
\begin{equation}
 \langle{\bf H}_{v}\rangle=(\omega-\lambda'v){\bf L}+
 \mu({\bf D}-v{\bf F})\,.
\end{equation}
We assume $\theta$ to be small and thus use
$\omega\approx\omega\cos2\theta$. We now have a system where the
effective vacuum oscillation frequencies for neutrinos are uniformly
distributed between $\omega\pm\lambda'$ and for antineutrinos between
$-\omega\pm\lambda'$. Even after removing the average common
precession of all modes, their evolution is still dominated by the
matter-flux term if $\lambda'\gg\mu$. In other words, collective
behavior now requires $\mu\agt\lambda'$ and not only $\mu\agt\omega$.

The simplest example is the ``flavor pendulum'' where for
$\epsilon=0$ and an isotropic neutrino gas one obtains the well-known
pendular motions of the polarization vectors. Matter does not disturb
this behavior, except that it takes logarithmically longer for the
motion to start~\cite{Hannestad:2006nj}. However, a matter flux, if
sufficiently strong, suppresses this motion and the polarization
vectors remain stuck to the ${\bf L}$ direction for both mass
hierarchies. If the neutrino distribution is not isotropic, the
ensemble quickly decoheres kinematically~\cite{Raffelt:2007yz}, an
effect that is also suppressed by a sufficiently strong matter flux.

We have verified these predictions in several numerical examples, but
have not explored systematically the transition between a ``weak''
and a ``strong'' matter flux because a homogeneous ensemble only
serves as a conceptual example where matter can have a strong
influence on self-induced transformations.

\section{Spherical Stream}                          \label{sec:stream}

The most general case of neutrino flavor evolution consists of an
ensemble evolving both in space and time. In practice, however, one
usually considers quasi-stationary situations where one asks for the
spatial flavor variation of a stationary neutrino flux streaming from
a source. The neutrino density decreases with distance so that one
can mimic this situation by a homogeneous system evolving in time
with a decreasing density, the expanding universe being a realistic
example. However, the analogy has important limitations because
collective oscillations introduce geometric complications into the
spatial-variation case.

The simplest nontrivial example is a perfectly spherical source (``SN
core'') that emits neutrinos and antineutrinos  like a
blackbody surface into space. The matter background is also taken to
be perfectly spherically symmetric, but of course varies with radius.
As a further simplification we consider monochromatic neutrinos and
antineutrinos that are all emitted with the same energy. Therefore,
the only variable necessary to classify the neutrino modes is their
angle of emission. The most useful variable is $u=\sin^2\vartheta_R$
where $\vartheta_R$ is the angle of emission at the neutrino sphere
at $r=R$ \cite{EstebanPretel:2007ec}. At a distance $r$ the radial
velocity of a mode $u$~is
\begin{equation}
v_{u,r}=\sqrt{1-u\,(R/r)^2}\,.
\end{equation}
For blackbody-like  emission the flux modes are uniformly
distributed in the interval $0\leq u\leq 1$.

The polarization vectors are taken to represent the neutrino flux
density because it is the flux integrated over a sphere of radius
$r$, not the density, that is conserved as a function of $r$.
Ignoring a possible matter flux, the EOMs
are~\cite{EstebanPretel:2007ec}
\begin{equation}
\partial_r{\bf P}_{u,r}={\bf H}_{u,r}\times{\bf P}_{u,r}\,,
\end{equation}
where the Hamiltonian is
\begin{equation}\label{eq:Ham1}
 {\bf H}_{u,r}=
 \frac{\omega {\bf B}+\lambda_r{\bf L}}{v_{u,r}}
 +\mu_r \left(\frac{{\bf D}_{r}}{v_{u,r}}-{\bf F}_{r}\right)\,.
\end{equation}
For antineutrinos we have, as always, $\omega\to-\omega$. Since the
polarization vectors describe the fluxes, the global density and flux
polarization vectors are
\begin{eqnarray}
 {\bf D}_r&=&\int_0^1\D u\,
 \frac{{\bf P}_{u,r}-\bar{\bf P}_{u,r}}{v_{u,r}}\,,
 \nonumber\\*
 {\bf F}_r&=&\int_0^1\D u\,
 \left({\bf P}_{u,r}-\bar{\bf P}_{u,r}\right)\,,
\end{eqnarray}
using the normalization $\int_0^1 \D u\,\bar{\bf P}_{u,r}=1$ and
$\int_0^1 \D u\,{\bf P}_{u,r}=1+\epsilon$. The matter coefficient
$\lambda_r=\sqrt2\,G_{\rm F}[n_{e^-}(r)-n_{e^+}(r)]$ encodes the
effective electron density at radius $r$ whereas
\begin{equation}
\mu_r=\mu_R\,\frac{R^2}{r^2}\,.
\end{equation}
Here, $\mu_R=\sqrt2 G_{\rm F} \Phi_{\bar\nu_e}(R)$ with
$\Phi_{\bar\nu_e}(R)$ the antineutrino flux at the neutrino sphere.
Therefore, $\mu_r$ always varies as $r^{-2}$ due to the geometric
flux dilution, whereas $\lambda_r$ is given by the detailed matter
profile of a SN model~\cite{Schirato:2002tg,Fogli:2003dw,Tomas:2004gr}.

If we ignore for the moment the neutrino-neutrino term in
Eq.~(\ref{eq:Ham1}), the EOM written in terms of the radial
coordinate $r$ simply re-parametrizes the distance along the
trajectory of a neutrino. Naturally, the oscillation pattern of a
trajectory that is tilted relative to the radial direction produces a
compressed oscillation pattern when projected on the radial
direction. In the absence of neutrino-neutrino interactions, this
simply means that along the radial direction the flavor variation of
the global neutrino stream decoheres kinematically, even if the
neutrinos are monochromatic and thus have the same vacuum oscillation
frequency along their trajectories.

In the presence of neutrino-neutrino interactions, kinematical
decoherence among different angular modes can still occur and in fact
is a self-accelerating process if the asymmetry $\epsilon$ is too
small. On the other hand, for a sufficient asymmetry, the radial
variation of different angular modes is collective and they behave
almost as if they were all emitted with the same angle relative to
the radial direction. While this quasi single-angle behavior is
theoretically not understood, numerically it has been consistently
observed~\cite{Duan:2006an, EstebanPretel:2007ec,Fogli:2007bk}.

The variation of the polarization vectors with the common radial
coordinate $r$ now acquires dynamical significance in that the
polarization vectors evolve differently than they would in the
absence of neutrino-neutrino interactions. From Eq.~(\ref{eq:Ham1})
it is obvious that the matter term is no longer the same for all
modes and thus can not be transformed away by going to a rotating
frame. This behavior does not depend on the radial variation of
$\lambda_r$---even a homogeneous medium would show this multi-angle
matter effect.

For quasi single-angle behavior to occur, $\epsilon$ must not be too
small, a condition that is probably satisfied in a realistic SN.
Therefore, the synchronization radius implied by Eq.~(\ref{eq:synch})
is always much larger than the neutrino-sphere radius $R$, allowing
us to expand the EOMs in powers of $R/r\ll1$. Using
\begin{equation}
v_{u,r}^{-1}=1+\frac{u}{2}\,\frac{R^2}{r^2}
\end{equation}
we find
\begin{equation}
 {\bf H}_{u,r}=(\omega {\bf B}+\lambda_r{\bf L})
 \left(1+\frac{u}{2}\,\frac{R^2}{r^2}\right)
 +\mu_r\,\frac{R^2}{2r^2}
 \left({\bf Q}_r+u{\bf F}_r\right)\,,
\end{equation}
where
\begin{equation}
{\bf Q}_r=\int_0^1\D u\,u
\left({\bf P}_{u,r}-\bar{\bf P}_{u,r}\right)
\end{equation}
and ${\bf F}_r$ is the same as before. At large $r$ a small
correction to $\omega$ is not crucial and can be ignored. The radial
variation of $\lambda_r$ is slow compared to the precession, so we
can go to a frame that rotates with a different frequency $\lambda_r$
at each radius. Finally the rotation-averaged Hamiltonian is
\begin{equation}
 \langle{\bf H}_{u,r}\rangle=(\omega +u\,\lambda_r^*)\,{\bf L}
 +\mu^*_r\,\left({\bf Q}_r+u{\bf F}_r\right)\,,
\end{equation}
where
\begin{eqnarray}\label{eq:mulambda}
 \lambda_r^*&=&\lambda_r\,\frac{R^2}{2r^2}\,,
 \nonumber\\*
 \mu_r^*&=&\mu_r\,\frac{R^2}{2r^2}=\mu_R\,\frac{R^4}{2r^4}
\end{eqnarray}
and we have assumed $\omega\cos2\theta\approx\omega$.

The multi-angle matter effect can be neglected if in the collective
region beyond the synchronization radius we have
\begin{equation}
\lambda^*_r\ll \mu_r^*
\end{equation}
equivalent to
\begin{equation}\label{eq:condition}
n_{e^-}-n_{e^+}\ll n_{\bar\nu_e}\,.
\end{equation}
In the opposite limit we expect that the large spread of effective
oscillation frequencies prevents collective oscillations. In this
case all polarization vectors remain pinned to the ${\bf L}$
direction and no flavor conversion occurs.

For intermediate values it is not obvious what will happen. One may
expect that the multi-angle matter effect triggers multi-angle
decoherence, destroying the quasi single-angle behavior. This indeed
occurs for the inverted hierarchy whereas in the normal hierarchy we
have not found any conditions where multi-angle decoherence was
triggered by the multi-angle matter effect. We recall that for a
sufficiently small $\epsilon$ multi-angle decoherence occurs even in
the normal hierarchy whereas no collective transformation arise for a
sufficiently large $\epsilon$~\cite{EstebanPretel:2007ec}.

\begin{figure}[ht]
\begin{center}
\epsfig{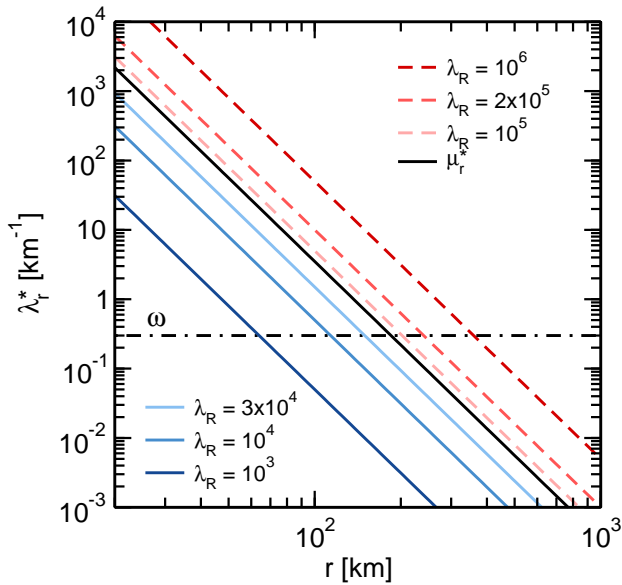}
\end{center}
\caption{Radial variation of $\mu^*_r$ and $\lambda_r^*$ for our
  numerical examples with the indicated value of $\lambda_R$. The
  value of $\omega$ considered in this work is also
  shown.} \label{fig:profiles}
\end{figure}

We illustrate these points with a numerical example where $R=10$~km,
$\omega=0.3~{\rm km}^{-1}$, $\theta=10^{-2}$, 
$\mu_R=7\times 10^4~{\rm km}^{-1}$ and
\begin{equation}
\lambda_r=\lambda_R\,\left(\frac{R}{r}\right)^n
\end{equation}
with $n=2$. This particular value of the power-law index leads to the
same radial dependence of $\mu^*_r$ and $\lambda_r^*$, see
Eq.~(\ref{eq:mulambda}).  In Fig.~\ref{fig:profiles} we show the radial
variation $\mu^*_r$ and $\lambda_r^*$ for different choices of
$\lambda_R$, between $10^3$~km$^{-1}$ and $10^6$~km$^{-1}$. Even for
the smallest matter effect, the ordinary MSW resonance, defined by the
condition $\lambda_r=\omega$, stays safely beyond the collective
region.

\begin{figure}
\begin{center}
\epsfig{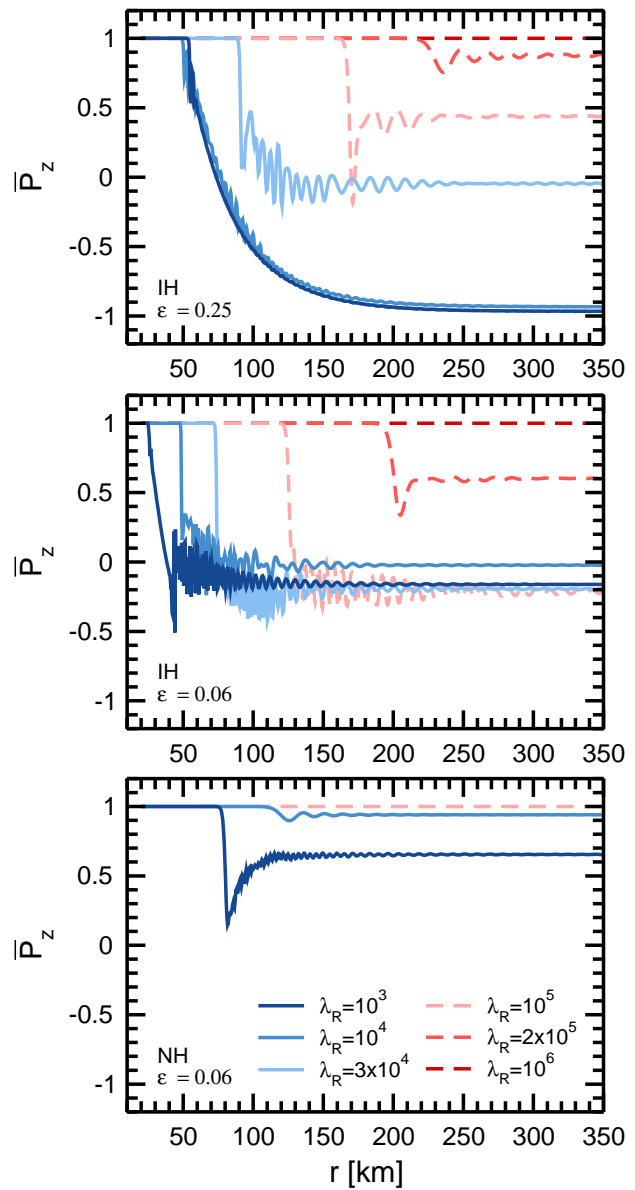}
\end{center}
\caption{Radial variation of $\bar P_z$ for three different scenarios:
  inverted mass hierarchy and $\epsilon = 0.25$ (top panel), inverted
  mass hierarchy and $\epsilon = 0.06$ (middle panel), and normal mass
  hierarchy and $\epsilon = 0.06$ (bottom panel).  In each 
  panel different values of $\lambda_R$ have been assumed, reported in the
  bottom panel.}
\label{fig:transformation}
\end{figure}

In Fig.~\ref{fig:transformation} we show the corresponding variation
of $\bar P_z$ for three different cases: inverted mass hierarchy and
$\epsilon = 0.25$ (top panel), inverted mass hierarchy and $\epsilon =
0.06$ (middle panel), and normal mass hierarchy and $\epsilon =
0.06$ (bottom panel).
In the top panel we observe the usual transformation for a small
matter effect, a complete suppression of transformations for a large
matter effect, and multi-angle decoherence for intermediate
cases. Repeating the same exercise for the normal mass hierarchy and
the same $\epsilon$ reveals no macroscopic influence of the matter
term.

For a sufficiently small $\epsilon$ one finds self-induced multi-angle
decoherence for both hierarchies. In the middle and bottom panels of
Fig.~\ref{fig:transformation} we show how with a sufficiently strong
matter effect the decoherence can be suppressed for both hierarchies.

\section{Discussion}                            \label{sec:discussion}

We have identified a new multi-angle effect in collective neutrino
transformations that is caused by a matter background. Previous
numerical studies of multi-angle effects had used a matter profile
that satisfies the condition Eq.~(\ref{eq:condition}) in the critical
region~\cite{Duan:2006an, Fogli:2007bk}. In other multi-angle studies
matter was entirely ignored~\cite{EstebanPretel:2007ec} and
otherwise, single-angle studies were performed. Therefore, the
multi-angle matter effect discussed here had escaped numerical
detection.

In many practical cases relevant for SN physics or in coalescing
neutron stars, the density of matter is probably small enough so that
this effect can be ignored. On the other hand, for iron-core SNe,
during the accretion phase the matter density can reach values and
can have such a profile that it is important. The possibility of such
large matter densities had led some of us to speculate that the
second-order difference between the $\nu_\mu$ and $\nu_\tau$
refractive effect could sometimes play an interesting role for
three-flavor collective transformations~\cite{EstebanPretel:2007yq}.
It is easy to show, however, that the density requirement for this
mu--tau effect to be important implies that the multi-angle matter
effect can not be avoided. In this sense, one complicated effect
caused by a large matter density annihilates another one.

If at early times the matter density profile is such that our
multi-angle effect is important, this will not be the case at later
times when the explosion has occurred and the matter profile
contracts toward the neutron star. In principle, therefore,
interesting time-dependent features in the oscillation probability
can occur.

A large matter effect can be ``rotated away'' from the EOMs when it is
identical for all modes. Here we have seen that even a perfectly
uniform medium provides a multi-angle variation of the matter effect.
We note that the matter fluxes would not be important, in contrast to
our first example of a homogeneous ensemble, because the relevant
quantity is the spread of the matter effect between different
modes. Therefore, whenever a flux term would be important, the matter
density term already provides a strong multi-angle effect.

Possible polarizations of the electron-positron background in the
strong SN magnetic fields would produce an additional axial term in
the matter potential, proportional to the scalar product of the
neutrino momentum with the electron polarization
direction~\cite{Nunokawa:1997dp}.  However, we expect that also this
effect should be negligible with respect to the matter density term.

In addition, the medium can have density variations caused by
convection and turbulence~\cite{Scheck:2007gw} that is known to affect
the MSW resonance under certain circumstances~\cite{Loreti:1995ae,
  Fogli:2006xy, Friedland:2006ta, Choubey:2007ga, Kneller:2007kg}.
Density variations in the transverse direction to the neutrino stream
lines may well cause important variations of the matter effect between
different modes. It remains to be investigated in which way collective
flavor transformations are affected.


\begin{acknowledgments}
We acknowledge partial support by the DFG (Germany) under grants TR-27
and SFB-676, by The Cluster of Excellence ``Origin and Structure of
the Universe,'' by the European Union under the ILIAS project
(contract No.\ RII3-CT-2004-506222) and an RT Network (contract No.\
MRTN-CT-2004-503369), and by the Spanish grant FPA2005-01269. A.E.\ is
supported by a FPU grant from the Spanish Government, A.M.\ by INFN
(Italy) through a post-doctoral fellowship, and P.S.\ by the US
Department of Energy and by NASA grant NAG5-10842. Fermilab is
operated by Fermi Research Alliance, LLC under Contract
No.~DE-AC02-07CH11359 with the United States Department of Energy.
\end{acknowledgments}


\end{document}